\documentclass{article}
\usepackage{spconf,amsmath,graphicx}

\usepackage{enumitem}
\usepackage{amsmath}
\usepackage{url}
\setlist{nosep, leftmargin=14pt}

\usepackage{mwe} 

\title{{Patch-Based Encoder-Decoder Architecture for Automatic Transmitted Light to Fluorescence Imaging Transition: Contribution to the LightMyCells Challenge}
\thanks{P\lowercase{ermission from \uppercase{IEEE} must be obtained for all other uses, in any current or future media, including reprinting/republishing this material for advertising or promotional purposes, creating new collective works, for resale or redistribution to servers or lists, or reuse of any copyrighted component of this work in other works.}}}
\name{Marek Wodzinski$^{1,2}$, Henning M\"{u}ller$^{2,3}$}
\address{$^1$Department of Measurement and Electronics, AGH University of Krakow, Krakow, Poland \\
$^2$Institute of Informatics, University of Applied Sciences Western Switzerland, Sierre, Switzerland \\
$^3$Medical Faculty, University of Geneva, Geneva, Switzerland
}

\begin{document}

%
\maketitle
\begin{abstract}
Automatic prediction of fluorescently labeled organelles from label-free transmitted light input images is an important, yet difficult task. The traditional way to obtain fluorescence images is related to performing biochemical labeling which is time-consuming and costly. Therefore, an automatic algorithm to perform the task based on the label-free transmitted light microscopy could be strongly beneficial. The importance of the task motivated researchers from the France-BioImaging to organize the LightMyCells challenge where the goal is to propose an algorithm that automatically predicts the fluorescently labeled nucleus, mitochondria, tubulin, and actin, based on the input consisting of bright field, phase contrast, or differential interference contrast microscopic images. In this work, we present the contribution of the AGHSSO team based on a carefully prepared and trained encoder-decoder deep neural network that achieves a considerable score in the challenge, being placed among the best-performing teams.
\end{abstract}
\section{Introduction}
\label{sec:intro}

The traditional way of obtaining fluorescence microscopy images is costly and time-consuming. It requires manual biochemical labeling that may be additionally perturbed by the exposure to excitation light and other factors. In contrast, label-free transmitted light microscopy images are non-invasive and easier to acquire. Therefore, an algorithm that could automatically predict the fluorescence microscopy images from modalities such as bright field (BF), phase contrast (PC), or differential interference contrast (DIC) could be beneficial to further speed up the research and practical applications of fluorescence imaging. The importance of the task motivated researchers from France-BioImaging to organize an open challenge during the IEEE ISBI 2024 conference, named LightMyCells~\cite{lmc_challenge}. The goal of the challenge is to automatically and accurately predict the fluorescence images based on the label-free transmitted light images, taking into account the high acquisition variability.

The current state-of-the-art in image-to-image tasks in fluorescence imaging consist of several notable contributions. For example, the DeepHCS~\cite{lee2021deephcs++} proposes a hybrid approach to perform the prediction based on an encoder-decoder framework with additional adversarial loss, achieving accurate and robust results. Another work applies a carefully tuned multi-head network performing multi-scale prediction~\cite{christiansen2018silico}. The most recent contribution compares the generative adversarial networks with gradient penalty to the traditional UNet architecture~\cite{cross2022label, ronneberger2015u}, presenting significant improvement of the adversarial learning compared to the traditional encoder-decoder architectures. Nevertheless, all the methods share a limitation that is addressed by the LightMyCells challenge - the works present a low diversity of data with limited applications, without an openly accessible database. In contrast, the LightMyCells challenge provides a heterogeneous dataset and open evaluation mechanism that allows researchers to reliably compare their contributions.

\begin{figure*}[!htb]
    \centering
    \includegraphics[width = 0.85\textwidth]{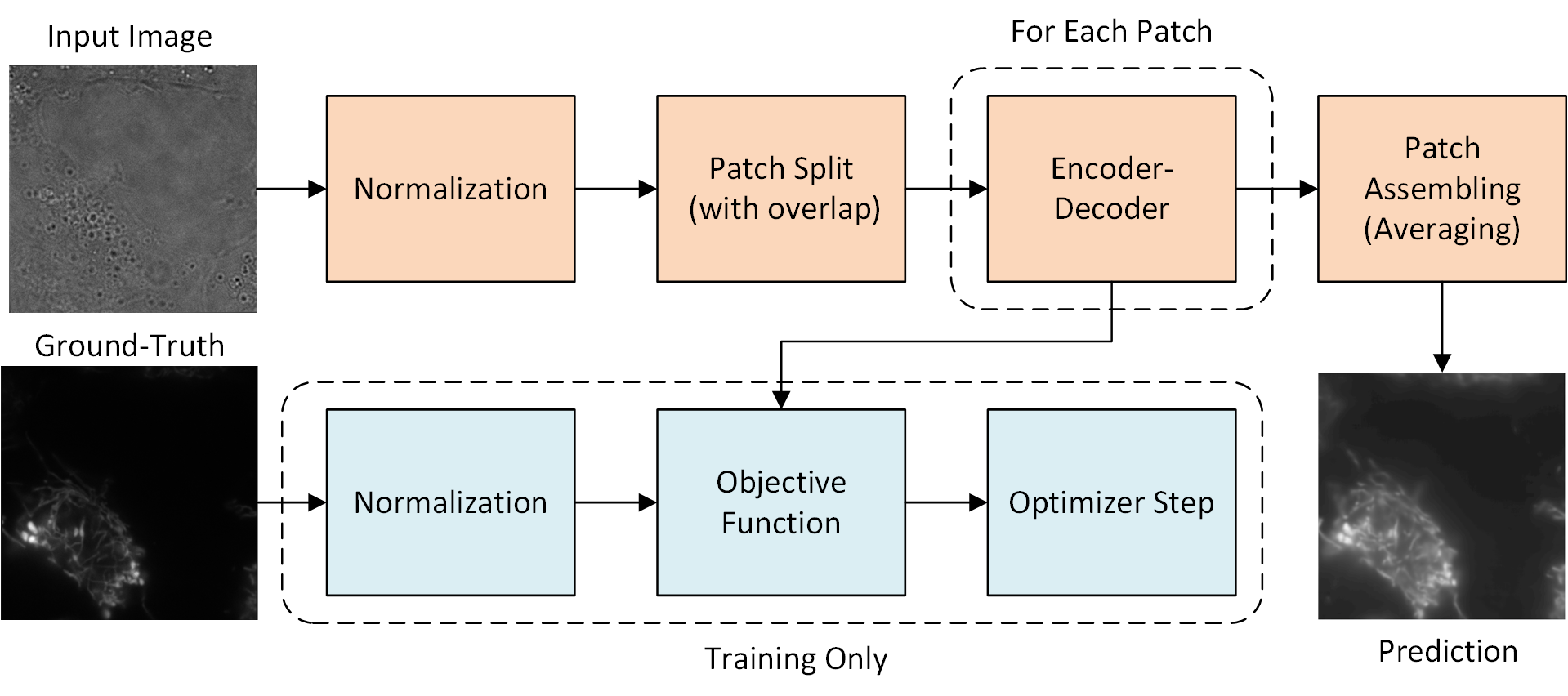}
    \caption{Visualization of the inference and training pipeline..}
    \label{fig:pipeline}
\end{figure*}

Several factors had to be taken into account when proposing a solution to the LightMyCells challenge. First, the training dataset is strongly heterogeneous, consisting of cases from different studies, imaging modalities, physical pixel sizes, resolutions, and zoom levels. Secondly, the data strongly suffers from the problem related to sparse annotations. The ground-truth, especially related to actin and tubulin, is not available for the majority of the cases. That makes it more difficult to propose and train a single model for all the organelles. Finally, the dataset is relatively big and the experiments require a significant amount of resources, even for lightweight models for which the input-output operations become the bottleneck. A successful contribution to the challenge has to take into account all the factors.

\textbf{Contribution: } We present our contribution to the LightMyCells challenge organized during the IEEE ISBI 2024 conference. Our contribution is based on the encoder-decoder network built on top of the RUNet network~\cite{wodzinski2023automatic}. We perform several ablation studies presenting comparisons between different architectures, objective functions, and training strategies. We show that, surprisingly, a method based on the simplest architecture and training strategy is the most effective. The proposed method scored 3rd place in the public leaderboard, however, the final challenge results are not available yet. We openly release both the source code, the Docker container used for the inference, and the final model~\cite{lmc_repository}.

\section{Methods}
\label{sec:methods}

\subsection{Overview}

The proposed method consists of the following steps: (i) preprocessing, (ii) patch-based prediction by encoder-decoder neural network, (iii) patch assembling and postprocessing. The overview is presented in Figure~\ref{fig:pipeline}.

\subsection{Preprocessing \& Postprocessing}

The input images are firstly normalized to [0-1] range and then split into fixed-size patches (512x512). During training, the ground-truth fluorescence images are also normalized to [0-1] range.

The output patches are assembled using the grid sampling and Hann-window-based averaging using the TorchIO library~\cite{torchio}. The output patches are not scaled or normalized in any other way.

\subsection{Encoder-Decoder Architecture}

The final model is based on the RUNet architecture~\cite{wodzinski2023automatic} that turned out to be successful in several other challenges. We compared the architecture to several other contributions like SwinUNETR~\cite{hatamizadeh2021swin}, UNETR~\cite{hatamizadeh2022unetr}, Attention UNet~\cite{oktay2018attention} available in the MONAI library~\cite{Cardoso_MONAI_An_open-source_2022}, however, it turned out that the RUNet architecture is the most effective (as shown in the ablation studies). The neural network takes as the input a single-channel transmitted light microscopy image and outputs a single-channel fluorescence image. The details related to the architecture are available in the associated repository~\cite{lmc_repository}.

\subsection{Training}

The model was trained using a fully supervised approach using an objective function being a linear combination of differentiable mean squared error (MSE), negative structural similarity index measure (SSIM), negative Pearson correlation coefficient (PCC), and cosine distance (CD):
\begin{equation}
\begin{split}
O(P, GT) = \alpha*MSE(P, GT) + \beta*SSIM(P, GT) \\ + \lambda*PCC(P, GT) + \omega*CD(P, GT),
\end{split}
\end{equation}
where $P$ is the predicted image, and $GT$ is the ground-truth image. The $\alpha$, $\beta$, $\lambda$, and $\omega$ were set to 1.0, 0.2, 0.1, and 0.1 respectively.

\begin{table*}[!htb]
\centering
\caption{Table presenting the quantitative results of the proposed method, together with ablation studies performed with respect to the training strategy, encoder-decoder architecture, patch size, and objective function. The significant differences between external and internal validation probably result from internal split at the image-level instead of the study-level. The best result for a particular ablation study is bolded.}
\renewcommand{\arraystretch}{1.0}
\resizebox{0.99\textwidth}{!}{%
\begin{tabular}{lcccccccccccccc}
\multicolumn{1}{c}{Method} & \multicolumn{5}{c}{Nucleus} & \multicolumn{5}{c}{Mitochondria} & \multicolumn{2}{c}{Tubulin} & \multicolumn{2}{c}{Actin} \tabularnewline
\hline
\multicolumn{1}{c}{}
& \multicolumn{1}{c}{MAE $\downarrow$} & \multicolumn{1}{c}{SSIM $\uparrow$} & \multicolumn{1}{c}{PCC $\uparrow$} & \multicolumn{1}{c}{CD $\downarrow$} & \multicolumn{1}{c}{ED $\downarrow$}
& \multicolumn{1}{c}{MAE $\downarrow$} & \multicolumn{1}{c}{SSIM $\uparrow$} & \multicolumn{1}{c}{PCC $\uparrow$} & \multicolumn{1}{c}{CD $\downarrow$} & \multicolumn{1}{c}{ED $\downarrow$}
& \multicolumn{1}{c}{SSIM $\uparrow$} & \multicolumn{1}{c}{PCC $\uparrow$}
& \multicolumn{1}{c}{SSIM $\uparrow$} & \multicolumn{1}{c}{PCC $\uparrow$}
\tabularnewline

\hline
\multicolumn{14}{c}{a) Final Results (External Validation - Grand-Challenge Platform)}
\tabularnewline

External Validation & 0.0700 & 0.7490 & 0.7750 & 0.1440 & 170.8290 & 0.1040 & 0.6530 & 0.5300 & 0.2390 & 222.4700 & 0.7200 & 0.6600 & 0.3770 & 0.0780
\tabularnewline

External Test & 0.0699 & 0.7466 & 0.7458 & 0.1693 & 176.6736 & 0.1032 & 0.6537 & 0.5999 & 0.2138 & 226.6678 & 0.6774 & 0.5642 & 0.5460 & 0.5995
\tabularnewline

\hline
\multicolumn{15}{c}{b) Training Strategy Ablation (512x512, RUNet, Combined Objective, Internal Validation)}
\tabularnewline

Separate-Encoder & \textbf{0.0572} & \textbf{0.7814} & \textbf{0.8129} & \textbf{0.1223} & \textbf{152.147} & \textbf{0.0921} & \textbf{0.7171} & \textbf{0.6228} & \textbf{0.1724} & \textbf{192.829} & \textbf{0.8114} & \textbf{0.8241} & \textbf{0.7829} & \textbf{0.8417}
\tabularnewline

Shared-Encoder & 0.0792 & 0.5955 & 0.6171 & 0.2718 & 217.991 & 0.1381 & 0.6492 & 0.5918 & 0.2291 & 239.949 & 0.6274 & 0.5972 & 0.6917 & 0.7174
\tabularnewline

\hline
\multicolumn{15}{c}{c) Deep Encoder-Decoder Architecture (512x512, Separate-Encoder, Combined Objective, Internal Validation)}
\tabularnewline

RUNet & \textbf{0.0572} & \textbf{0.7814} & \textbf{0.8129} & \textbf{0.1223} & \textbf{152.147} & \textbf{0.0921} & \textbf{0.7171} & \textbf{0.6228} & \textbf{0.1724} & \textbf{192.829} & \textbf{0.8114} & \textbf{0.8241} & \textbf{0.7829} & \textbf{0.8417}
\tabularnewline

UNETR & 0.0673 & 0.7299 & 0.7619 & 0.1329 & 172.492 & 0.1092 & 0.6814 & 0.5714 & 0.1984 & 203.875 & 0.7291 & 0.7418 & 0.5914 & 0.5378
\tabularnewline

SwinUNETR & 0.0729 & 0.7172 & 0.7589 & 0.1341 & 178.941 & 0.1121 & 0.6785 & 0.5682 & 0.1992 & 207.869 & 0.6814 & 0.7195 & 0.5589 & 0.5129
\tabularnewline

AttentionUNet & 0.0643 & 0.7581 & 0.7917 & 0.1281 & 163.952 & 0.0973 & 0.7074 & 0.6192 & 0.1814 & 195.914 & 0.7615 & 0.7914 & 0.7458 & 0.7694
\tabularnewline
\hline
\multicolumn{15}{c}{d) Patch-Size (Sperate-Encoder, RUNet, Combined Objective, Internal Validation)}
\tabularnewline
512x512 & \textbf{0.0572} & \textbf{0.7814} & \textbf{0.8129} & \textbf{0.1223} & \textbf{152.147} & \textbf{0.0921} & \textbf{0.7171} & \textbf{0.6228} & \textbf{0.1724} & \textbf{192.829} & \textbf{0.8114} & \textbf{0.8241} & \textbf{0.7829} & \textbf{0.8417}
\tabularnewline
256x256 & 0.0914 & 0.6129 & 0.7149 & 0.1791 & 219.978 & 0.1381 & 0.6549 & 0.5414 & 0.2192 & 252.427 & 0.7219 & 0.7791 & 0.6917 & 0.7591
\tabularnewline
128x128 & 0.1274 & 0.4721 & 0.6241 & 0.2792 & 298.591 & 0.2081 & 0.5281 & 0.4817 & 0.2914 & 334.891 & 0.5914 & 0.6191 & 0.4591 & 0.5258
\tabularnewline
Resampling (1024x1024) & 0.0949 & 0.5919 & 0.6919 & 0.1892 & 227.821 & 0.1341 & 0.6626 & 0.5519 & 0.2184 & 247.329 & 0.6289 & 0.6596 & 0.2291 & 0.3347
\tabularnewline
\hline
\multicolumn{15}{c}{e) Objective Function (Sperate-Encoder, RUNet, 512x512, Internal Validation)}
\tabularnewline
Combined Objective & 0.0572 & \textbf{0.7814} & 0.8129 & \textbf{0.1223} & 152.147 & 0.0921 & 0.7171 & 0.6228 & \textbf{0.1724} & 192.829 & \textbf{0.8114} & \textbf{0.8241} & \textbf{0.7829} & \textbf{0.8417}
\tabularnewline
MSE & \textbf{0.0551} & 0.7219 & 0.7419 & 0.1591 & \textbf{149.544} & \textbf{0.0874} & 0.6342 & 0.5549 & 0.1945 & \textbf{188.851} & 0.6182 & 0.6814 & 0.5182 & 0.5749
\tabularnewline
SSIM & 0.0814 & 0.7782 & 0.7519 & 0.1492 & 202.507 & 0.1371 & \textbf{0.7232} & 0.5319 & 0.2119 & 236.421 & 0.7927 & 0.7329 & 0.7691 & 0.7572
\tabularnewline
PCC & 0.0729 & 0.7491 & \textbf{0.8251} & 0.1417 & 183.271 & 0.1271 & 0.6728 & \textbf{0.6459} & 0.2051 & 221.819 & 0.7417 & 0.8217 & 0.7291 & 0.8126
\tabularnewline
\hline
\label{tab:results}
\end{tabular}}
\end{table*}

\subsection{Experimental Setup}

The internal experiments were performed using 80\% and 20\% split ratios (training/internal validation). The usage of 5-fold cross-validation was rejected due to the computational complexity of the training procedure. All the experiments were performed using PyTorch and PyTorch lightning libraries~\cite{falcon2019pytorch} using a single NVIDIA A100 GPU (40GB). 

We performed ablation studies concerning: (i) training strategy (separate encoder-decoder network for each organelle vs shared encoder), (ii) encoder-decoder architecture, (iii) patch-size, and (iv) objective functions. The nucleus, mitochondria, and tubulin images were augmented by random flips across both axes. The actin model (due to the low amount of data) was additionally augmented by random elastic transforms. Random elastic transforms for nucleus, mitochondria, and tubulin images resulted in an unacceptable increase in the training time.

\section{Results}
\label{sec:results}

We evaluate the image-to-image translation quality by using mean absolute error (MAE), SSIM, PCC, CD, and Euclidean distance (ED) for mitochondria and nucleus, and SSIM and PCC for tubulin and actin, following the conventions from the challenge organizers. The evaluation is performed separately for all organelles, however, we do not evaluate the quality with respect to the deviation from the best zoom level. Each time a significant improvement is reported, it means that the claim is supported by the Wilcoxon signed-rank test with a p-value lower than 0.01.

The results for the external (Grand-Challenge-based) validation and test sets are reported in Table~\ref{tab:results}a). The results are comparable to other challenge participants, however the final challenge results are not available at the time of writing the paper.

The ablation study related to the influence of training strategy is presented in Table~\ref{tab:results}b). The study confirms that the use of separate encoder-decoder architecture results in better performance than the shared encoder approach, even though the shared encoder has more than twice as many parameters as the final solution.

The experiments related to the influence of encoder-decoder architecture are presented in Table~\ref{tab:results}c). The Table presents significant improvement by using the traditional RUNet architecture compared to UNETR, SwinUNETR, or Attention UNet. It confirms that for such tasks the data preparation and training strategy are more influential factors than the most recent network architectures. 

The influence of patch size is reported in Table~\ref{tab:results}d). It can be noted that small patch sizes (128x128, 256x256) result in significantly worse results compared to the larger patch size (512x512). On the other hand, resampling the images to a fixed resolution (1024x1024) results in worse performance compared to the optimized patch-based approach.

Finally, the effect of combining the objective function is presented in Table~\ref{tab:results}e). It can be seen that the usage of combined objective function overall improves the average performance, and for SSIM and PCC sometimes even improves the absolute outcomes. It suggests that the use of the combined objective function somehow improves generalizability or decreases the risk of reaching local minima.

\section{Discussion}
\label{sec:discussion}

The proposed method achieves results comparable to the best-performing challenge participants. The most significant improvements are related to the use of a patch-based approach with appropriate patch size, the use of separate models for each organelle, and combining different objective functions during training.

The results of using an architecture with shared encoder and separate decoders for each organelle turned out to be significantly worse (p-value $<$ 0.01) than separate encoder-decoder architectures. Probably a further division into different types of imaging modalities (BF, DIC, PC) could further increase the accuracy, however, such an approach could lead to the problem of significant overfitting due to the limited amount of training data, especially for actin and tubulin. 

Another important point is the necessity to introduce dedicated study-based sampling of the data during training. The majority of the training data represents just a few studies. Random sampling of the images results in overfitting the model to a particular study with limited generalizability to the less represented studies.

The use of the patch-based approach significantly improves the results compared to direct resampling (p-value $<$ 0.01). Resampling the images into single resolution decreases the ability to recover small details and results in instant overfitting because the patch-based approach serves as a natural augmentation strategy. However, the patch-based should be large enough, too small patches result in worse outcomes.

Importantly, the linear combination of different objective functions (MSE, SSIM, PC, CD) improves the results when compared to the functions used in separation. It stabilizes the network training and makes it more resistant to local minima.

In future work, we will consider extending the network by utilizing metadata in the training and inference processes. Unfortunately, we could not check our ideas related to the metadata directly during the competition due to significant time restrictions. Nevertheless, we forecast that inference-time regularization or prompting related to the metadata of the input images (physical pixel size, imaging modality) could significantly improve the method performance, especially for less represented cases like actin in DIC or tubulin in BF.

To conclude, the proposed method achieved considerable scores in the competition, even though we used a relatively simple approach and training strategy. The use of separate models for each organelle improved the results compared to a more universal network attempting to address the problem of sparse annotations. Nevertheless, we assume that a smart mechanism to incorporate the metadata information during training and inference could further improve the results.

\section{Compliance with ethical standards}
\label{sec:ethics}

This research study was conducted retrospectively using microscopy data made available in open access by LightMyCells Challenge organizers. Ethical approval was not required as confirmed by the license attached with the open-access data.

\section{Acknowledgments}
\label{sec:acknowledgments}

The research was partially supported by the program "Excellence Initiative - Research University" for AGH University. We gratefully acknowledge Polish high-performance computing infrastructure PLGrid (HPC Center: ACK Cyfronet AGH) for providing computer facilities and support within computational grant no. PLG/2024/017079.

\bibliographystyle{IEEEbib}
\bibliography{refs}

\end{document}